\documentclass[aps,prd,showpacs,onecolumn,preprintnumbers,notitlepage,
nofootinbib,tightenlines,10pt]{revtex4-1}
\usepackage{amsmath,bm}
\usepackage{times}
\usepackage{amsfonts}
\usepackage{amssymb}
\usepackage{braket}
\usepackage{color,graphicx}
\usepackage{epstopdf}
\usepackage{latexsym}
\usepackage[dvipsnames]{xcolor}
\usepackage{slashed}
\usepackage{CJK,upgreek,fancyhdr}
\usepackage{float}
\usepackage{multirow}
\usepackage{hyperref}
\hypersetup{colorlinks,citecolor=nicegreen,linkcolor=nicered,urlcolor=RoyalBlue}
\usepackage{enumitem}
\usepackage{natbib}
\usepackage{relsize}
\usepackage[left=2.5cm,right=2.5cm,top=2.0cm,bottom=2.5cm]{geometry}
\newcommand{\beq}{\begin{eqnarray}}
\newcommand{\eeq}{\end{eqnarray}}
\newcommand{\non}{\nonumber\\ }

\newcommand{\psl}{ P \hspace{-2.4truemm}/ }

\newcommand{\epsl}{\epsilon \hspace{-1.6truemm}/\,  }

\def\lsim{ {\ \lower-1.2pt\vbox{\hbox{\rlap{$<$}\lower6pt\vbox{\hbox{$\sim$}
}}}\ } }
\def\gsim{ {\ \lower-1.2pt\vbox{\hbox{\rlap{$>$}\lower6pt\vbox{\hbox{$\sim$}
}}}\ } }

\def \jhep{ J. High Energy Phys.  }

\definecolor{Red}{rgb}{1.,0.,0.}
\definecolor{Blue}{rgb}{0.,0.,1.}
\definecolor{RoyalBlue}{rgb}{0.0, 0.14, 0.4}
\definecolor{nicered}{rgb}{0.7,0.1,0.2}
\definecolor{nicegreen}{rgb}{0.1,0.4,0.2}

\def\orcid#1{\kern .08em\href{https://orcid.org/#1}{\includegraphics[keepaspectratio,width=0.75em]{ORCID_iD.png}}}

\begin{document}
\begin{CJK*}{GB}{gbsn}
\title{\boldmath
Study of $B_c \to \chi_{cJ}\ (P, V)$ decays in the improved perturbative QCD formalism}
\author{Xin~Liu\orcid{0000-0001-9419-7462}}
\affiliation{Department of Physics,
Jiangsu Normal University, Xuzhou 221116, People's Republic of China}

\author{Hsiang-nan~Li\orcid{0000-0003-4080-4022}}
\affiliation{ Institute of Physics, Academia Sinica, Taipei, Taiwan 115, Republic of China}

\author{Zhen-Jun~Xiao\orcid{0000-0002-4879-209X}}
\affiliation{Department of Physics and Institute of Theoretical Physics,\\
	Nanjing Normal University, Nanjing 210023, People's Republic of China}


\date{\today{}}

\begin{abstract}

Motivated by the recent LHCb measurements of the ratios between the branching ratios (BRs),
$R_{\chi_{c2}/J/\psi} \equiv {\rm BR}(B_c^+ \to \chi_{c2} \pi^+)
/{\rm BR}(B_c^+ \to J/\psi \pi^+)$ and $R_{\chi_{c1}/\chi_{c2}} \equiv {\rm BR}(B_c^+
\to \chi_{c1} \pi^+)/{\rm BR}(B_c^+ \to \chi_{c2} \pi^+)$, we analyze the decays
$B_c^+ \to \chi_{cJ} (P, V)^+$ in the improved perturbative QCD (iPQCD) formalism, where
$\chi_{cJ}$ denotes the $p$-wave charmonia with $J=0, 1, 2$, and $P (V)$ denotes the
pseudoscalars $\pi$ and $K$ (the vectors $\rho$ and $K^*$). Our results $R_{\chi_{c2}/J/\psi}
= 0.32 \pm 0.05$ and $R_{\chi_{c1}/\chi_{c2}}= 0.11 \pm 0.01$ at leading
order in the strong coupling $\alpha_s$ agree well with the data within uncertainties.
The result $R_{\chi_{c0}/J/\psi} \equiv {\rm BR}(B_c^+ \to \chi_{c0} \pi^+)
/{\rm BR}(B_c^+ \to J/\psi \pi^+) =  0.17 \pm 0.02$ is lower than
$1.41^{+0.50}_{-0.45}$ inferred by the $B_c^+ \to \chi_{c0} \pi^+$ measurement. The
large longitudinal polarization fractions imply that these components dominate the
$B_c^+ \to \chi_{c1,c2} \rho^+$ BRs. A peculiar constructive (destructive) interference
between the twist-2 and twist-3 contributions is identified in the
$B_c^+\to \chi_{c1} \rho^+$ ($B_c^+\to \chi_{c1} \pi^+$) mode, but not in the
corresponding ones with $\chi_{c0,c2}$ mesons. Inputting the BRs of the strong decays
$\chi_{c0,c2} \to \pi^+ \pi^- / K^+ K^-$ and $\chi_{cJ} \to \pi^+\pi^-(\pi^+ \pi^-/K^+K^-)$
reported by the BESIII Collaboration, we obtain those of $B_c^+ \to \chi_{cJ} (\rho, \pi)^+$
followed by secondary decay chains under the narrow-width approximation. All the above iPQCD
predictions can be confronted by more precise experiments in the future to deepen our
understanding of QCD dynamics in $B_c\to\chi_{cJ}$ transitions.

\end{abstract}

\maketitle


\newpage
%
%

\section{Introduction}\label{sec:Intro}

A $B_c$ meson is the ground state of a unique meson family consisting of two different heavy
flavors, i.e., $b$ and $c$ quarks~\cite{Brambilla:2004wf,Brambilla:2010cs}. It is also the
only meson whose constituent decays compete with each other. Undoubtedly, the rich and
complicated dynamics in $B_c$ meson decays, being disparate from those in charmonia
and bottomonia, deserves thorough studies. On the experimental side, the LHCb has
accomplished the first upgrade of its detector and resumed running since
2022~\cite{Gao:2024lhc}. A huge amount of data with superior quality will offer a precious
opportunity for analyzing the production and decay of $B_c$ mesons and for promoting
the $B_c$ physics into a precision era. On the theoretical side, the dramatically distinct
predictions for $B_c$ meson decays in the literature indicate that our understanding
of the involved dynamics is still far from complete. More endeavors need to be devoted
to this field. The present work will concentrate on the decays $B_c^+ \to \chi_{cJ} (P, V)^+$,
where $P (V)$ denotes the pseudoscalars $\pi$ and $K$ (the vectors $\rho$ and $K^*$).
The $p$-wave charmonia $\chi_{cJ}$ with $J=0, 1, 2$ represent the scalar $\chi_{c0}(1P)$,
the axial-vector $\chi_{c1}(1P)$ and the tensor $\chi_{c2}(1P)$.

Recently, the $B_c^+ \to \chi_{c2} \pi^+$ decay was observed for the first time by the
LHCb with a significance exceeding seven standard deviations. The relative ratio between
the $B_c^+ \to \chi_{c2} \pi^+$ and $B_c^+ \to J/\psi \pi^+$ branching ratios (BRs)
was measured as~\cite{LHCb:2023fqn}
\beq
R^{\rm Exp}_{\chi_{c2}/J/\psi} &\equiv&
\frac{{\rm BR}(B_c^+ \to \chi_{c2} \pi^+)}
{{\rm BR}(B_c^+ \to J/\psi \pi^+)} =
0.37 \pm 0.06 \pm 0.02 \pm 0.01\;,
\label{eq:r-c2oj-e}
\eeq
where the first uncertainty is statistical, the second is systematic, and the third is
associated with the $\chi_{c2} \to J/\psi \gamma$ BR. As to the decay
$B_c^+ \to \chi_{c1} \pi^+$, the LHCb set the upper limit of the ratio between the
$B_c^+ \to \chi_{c1}\pi^+$ and $B_c^+ \to \chi_{c2} \pi^+$ BRs at the $90\%$
confidence level~\cite{LHCb:2023fqn},
\beq
R^{\rm Exp}_{\chi_{c1}/\chi_{c2}} &\equiv&
\frac{{\rm BR}(B_c^+ \to \chi_{c1} \pi^+)}
{{\rm BR}(B_c^+ \to \chi_{c2} \pi^+)}
< 0.49\;.
\label{eq:r-c1oc2-e}
\eeq
The LHCb also reported the evidence for the $B_c^+ \to \chi_{c0} \pi^+$ mode at a significance
of four standard deviations~\cite{LHCb:2016utz},
\beq
\frac{\sigma(B_c^+)}{\sigma(B^+)} \times {\rm BR}(B_c^+ \to \chi_{c0} \pi^+) &=&
(9.8^{+3.4}_{-3.0}({\rm stat}) \pm 0.8({\rm syst}))  \times 10^{-6}\;,
\label{eq:BR-c0pi}
\eeq
with the $B_c^+$ and $B^+$ meson production cross sections, $\sigma(B_c^+)$ and $\sigma(B^+)$,
respectively. If the ratio $\sigma(B_c^+)/\sigma(B^+)$ is determined definitely, the absolute
$B_c^+ \to \chi_{c0} \pi^+$ BR can be deduced, and vice versa.

\begin{table}[hbt]
\caption{The $B_c^+ \to \chi_{c2} \pi^+$ BRs and the ratios
between the $B_c^+ \to \chi_{c2} \pi^+$ and $B_c^+ \to J/\psi \pi^+$ BRs from various
theoretical approaches and the data. }
\label{tab:Rs-c2}
\begin{center}\vspace{-0.3cm}{\tiny
\begin{tabular}[t]{c|c|c|c|c|c|c|c|c|c|c|c|c|c}
\hline\hline
& Data~\cite{LHCb:2023fqn}
 &\cite{Chang:2001pm}
 &\cite{Kiselev:2001zb}
 &\cite{LopezCastro:2002woj}
 &\cite{Ivanov:2006ni}
 &\cite{Hernandez:2006gt}
 &\cite{Ebert:2010zu}
 &\cite{Wang:2011jt}
 &\cite{Rui:2017pre}
 &\cite{Zhu:2017lwi}
 \footnote{
These results are from the leading-order nonrelativistic QCD (NRQCD)
with relativistic corrections.}
 & \cite{Chen:2021vmb}
 \footnote{
 These results are from the next-to-leading-order NRQCD. The corresponding
 leading-order BR is given by $(9.7^{+14.0}_{-4.5}) \times 10^{-3}$.}
 & \cite{Losacco:2023uvp}
 &\cite{Lu:2025bvi}
 \footnote{
 The values in the parentheses were obtained without
 Coulomb-like corrections in the same framework based on three-point QCD sum rules.}
 \\
\hline
$10^{4}\cdot{\rm BR}(B_c^+ \to \chi_{c2} \pi^+)$
& $-$
& 2.5& 79.0& 0.75 & 4.6 & 2.2  & 3.8  & $2.1 \pm 0.5$ &$40.0^{+10.0}_{-9.4}$&$43.7^{+22.3}_{-15.2}$& $240^{+310}_{-110}$  & $0.37^{+0.14}_{-0.12}$
& $130.3^{+31.7}_{-5.1} (16.6^{+4.2}_{-1.2})$
\\
\hline
$R_{\chi_{c2}/{J/\psi}}\equiv \frac{{\rm BR}(B_c \to \chi_{c2} \pi)}{{\rm BR}(B_c \to J/\psi \pi)}$
& $0.37 \pm 0.06$
&0.14& 1.37& $0.069$ & 0.27 & 0.29 & 0.62 & $0.19^{+0.03}_{-0.03}$&$1.72^{+0.06}_{-0.13}$ & $1.50^{+0.48}_{-0.32}$
& $8.25^{+8.22}_{-2.86}$
& $0.049^{+0.014}_{-0.015}$
& $5.21^{+1.38}_{-0.00} (6.15^{+1.55}_{-0.21})$
\\
\hline \hline
\end{tabular}}
\end{center}
\end{table}

\begin{table}[hbt]
\caption{ Same as Table~\ref{tab:Rs-c2} but for the decay $B_c^+ \to \chi_{c1} \pi^+$.
}
\label{tab:Rs-c1}
\begin{center}\vspace{-0.3cm}{\tiny
\begin{tabular}[t]{c|c|c|c|c|c|c|c|c|c|c|c|c}
\hline\hline
& Data~\cite{LHCb:2023fqn}
 &\cite{Chang:2001pm}
 &\cite{Kiselev:2001zb}
 &\cite{Ivanov:2006ni}
 &\cite{Hernandez:2006gt}
 &\cite{Ebert:2010zu}
 &\cite{Wang:2011jt}
 &\cite{Rui:2017pre}
 &\cite{Zhu:2017lwi}
 & \cite{Chen:2021vmb}
 &\cite{Lu:2025bvi}
 &\cite{Zhang:2023ypl}
 \\
\hline
$10^{4}\cdot{\rm BR}(B_c^+ \to \chi_{c1} \pi^+)$
& $-$
& 0.7& 0.89 & 0.68 & 0.014  & 2.0  & $0.21 \pm 0.02$ &$5.1^{+1.1}_{-1.2}$&$0.64^{+0.11}_{-0.16}$& $2.3^{+2.8}_{-1.0}$
& $0.70^{+0.25}_{-0.26} (0.11^{+0.05}_{-0.04})$
&$1.3$
\\
\hline
$R_{\chi_{c1}/{\chi_{c2}}}\equiv \frac{{\rm BR}(B_c^+ \to \chi_{c1} \pi^+)}{{\rm BR}(B_c^+ \to \chi_{c2} \pi^+)}$
& $ < 0.49$
&0.28& 0.01 & 0.15 & 0.01 & 0.53 & $0.10^{+0.02}_{-0.01}$&$0.13^{+0.00}_{-0.01}$ & $0.02$
& $0.01$
& $0.005 (0.007)$
& $-$
\\
\hline
$R_{\chi_{c1}/{J/\psi}}\equiv \frac{{\rm BR}(B_c^+ \to \chi_{c1} \pi^+)}{{\rm BR}(B_c^+ \to J/\psi \pi^+)}$
& $-$
&0.04& 0.02 & 0.04 & 0.002 & 0.33 & $0.02$&$0.22^{+0.01}_{-0.02}$ & $0.02$
& $0.08^{+0.07}_{-0.03}$
& $0.03^{+0.00}_{-0.01} (0.04^{+0.01}_{-0.00})$
& $0.07^{+0.01}_{-0.01}$
\\
\hline \hline
\end{tabular}}
\end{center}
\end{table}

The decays $B_c^+ \to \chi_{cJ}\pi^+$ with the angular momenta $J=0,1,2$ have been
investigated in various theoretical frameworks~\cite{Chang:2001pm,Kiselev:2001zb,
LopezCastro:2002woj,Ivanov:2006ni,Hernandez:2006gt,Ebert:2010zu,Wang:2011jt,Rui:2017pre,
Zhu:2017lwi,Chen:2021vmb,Losacco:2023uvp}, and different predictions for their BRs and
relative ratios are presented in the Tables~\ref{tab:Rs-c2}, ~\ref{tab:Rs-c1},
and~\ref{tab:Rs-c0}. It has been a consensus that the two larger
$B_c^+ \to (\chi_{c0,c2}) \pi^+$ BRs
are of similar order of magnitude, and the $B_c^+ \to \chi_{c1} \pi^+$ BR is lower.
Despite this qualitative observation, the tables exhibit obvious quantitative
discrepancies among the predictions. Big spreads of the results for
${\rm BR}(B_c^+ \to \chi_{c2}\pi^+)$ and for $R_{\chi_{c2}/J/\psi}$ over three orders of
magnitude are seen in Table~\ref{tab:Rs-c2}. Only the two ratios $R_{\chi_{c2}/J/\psi}$
from~\cite{Ivanov:2006ni,Hernandez:2006gt} are roughly compatible with the recent LHCb data,
though the predicted ${\rm BR}(B_c^+ \to \chi_{c2} \pi^+)$ in Refs.~\cite{Chang:2001pm,
Ivanov:2006ni,Hernandez:2006gt,Ebert:2010zu,Wang:2011jt} are close to each other.
A broad range of ${\rm BR}(B_c^+ \to \chi_{c1} \pi^+)$ also
exists in Table~\ref{tab:Rs-c1}. Nevertheless, all $R_{\chi_{c1}/\chi_{c2}}$
except the one in~\cite{Ebert:2010zu} meet the upper bound set by the LHCb.
Almost all the predicted ${\rm BR}(B_c^+ \to \chi_{c0} \pi^+)$ in Table~\ref{tab:Rs-c0}
except the one in~\cite{Losacco:2023uvp} are of ${\cal O}(10^{-4})$ or higher.
These BRs vary significantly, and, in particular, the value from the NRQCD
factorization~\cite{Chen:2021vmb} manifests huge next-to-leading order corrections.
The notable disparity summarized above strongly suggests the necessity of
further explorations into this challenging subject in alternative approaches.

\begin{table}[hbt]
\caption{ Same as Table~\ref{tab:Rs-c2} but for the decay $B_c^+ \to \chi_{c0} \pi^+$.
}
\label{tab:Rs-c0}
\begin{center}\vspace{-0.3cm}{\tiny
\begin{tabular}[t]{c|c|c|c|c|c|c|c|c|c|c|c|c|c}
\hline\hline
& Data
 &\cite{Chang:2001pm}
 &\cite{Kiselev:2001zb}
 &\cite{Ivanov:2006ni}
 &\cite{Hernandez:2006gt}
 &\cite{Ebert:2010zu}
 &\cite{Wang:2011jt}
 &\cite{Rui:2017pre}
 &\cite{Zhu:2017lwi}
 & \cite{Chen:2021vmb}
 & \cite{Losacco:2023uvp}
 &\cite{Lu:2025bvi}
 &\cite{Zhang:2023ypl}
 \\
\hline
$10^{4}\cdot{\rm BR}(B_c^+ \to \chi_{c0} \pi^+)$
& $-$
& 2.8& 98.0 & 5.5 & 2.6  & 2.1 & $3.1 \pm 0.4$ &$16.0^{+3.6}_{-3.7}$&$64.7^{+18.6}_{-13.2}$& $280^{+380}_{-130}$  & $0.24^{+0.09}_{-0.08}$
& $19.6^{+2.1}_{-3.0} (2.4^{+0.5}_{-0.4})$
& $6.6^{+0.8}_{-0.7}$
\\
\hline
$R_{\chi_{c0}/{\chi_{c2}}}\equiv \frac{{\rm BR}(B_c^+ \to \chi_{c0} \pi^+)}{{\rm BR}(B_c^+ \to \chi_{c2} \pi^+)}$
& $ -$
&1.12& 1.24 & 1.20 & 1.18 & 0.55 & $1.48^{+0.21}_{-0.13}$&$0.40^{+0.01}_{-0.00}$ & $1.48^{+0.33}_{-0.22}$
& $1.17^{+0.03}_{-0.02}$
& $0.658$
& $0.15^{+0.00}_{-0.02} (0.14^{+0.00}_{-0.01})$
& $-$
\\
\hline
$R_{\chi_{c0}/{J/\psi}}\equiv \frac{{\rm BR}(B_c^+ \to \chi_{c0} \pi^+)}{{\rm BR}(B_c^+ \to J/\psi \pi^+)}$
& $-$
&0.16& 1.70 & 0.32 & 0.34 & 0.34 & $0.28 \pm 0.01$&$0.69^{+0.03}_{-0.07}$ & $2.22^{+0.27}_{-0.08}$
& $9.62^{+10.14}_{-3.40}$
& $0.032$
& $0.78^{+0.09}_{-0.08} (0.89^{+0.11}_{-0.06})$
& $0.34$
\\
\hline \hline
\end{tabular}}
\end{center}
\end{table}

We shall analyze the $B_c^+ \to \chi_{cJ} (P, V)^+$ decays in the improved perturbative
QCD (iPQCD) formalism, which incorporates the charm quark mass effects into the newly
established Sudakov resummation~\cite{Liu:2018kuo,Liu:2020upy,Liu:2023kxr},
besides keeping them in the hard decay kernels as done in the conventional
PQCD calculations~\cite{Xiao:2013lia}. This framework with the consistent treatment
of the charm quark effects at leading order in the strong coupling $\alpha_s$ is thus
expected to refine the previous PQCD studies on hadronic $B_c$ meson decays. It has been
demonstrated that the iPQCD predictions for the $B_c^+ \to J/\psi M^+$ modes, which are
dominated by factorizable emission contributions, agree well with the available
data~\cite{Liu:2023kxr}. It will be shown that the relative ratios among the
$B_c^+ \to (J/\psi, \chi_{c1,c2}) \pi^+$ BRs from the iPQCD also match the LHCb
measurements. The longitudinal polarization fractions for the $B_c^+\to\chi_{c1,c2}\rho^+$
decays are computed, and found to be nearly unity. Inputting the BRs of the strong
decays $\chi_{c0,c2} \to \pi^+ \pi^- / K^+ K^-$~\cite{BESIII:2025tho} and
$\chi_{cJ} \to \pi^+\pi^-(\pi^+ \pi^-/K^+K^-)$~\cite{BESIII:2024jmr,ParticleDataGroup:2024cfk},
we estimate those of multibody $B_c$ meson decays via charmonium resonances under the
narrow-width approximation. This work justifies the reliability of our theoretical setup
for $B_c$ meson decays, explores the involved QCD dynamics, and helps probe the inner
structure of charmonium states.

The rest of the paper is organized as follows. We present the iPQCD factorization
formulas in Sec.~\ref{sec:form}, which convolute the perturbative $b$ quark decay
kernels with the nonperturbative wave functions for the initial and final
states in parton transverse momenta~\cite{Liu:2018kuo}. Numerical outcomes and
related phenomenological insights are discussed and elaborated in Sec.~\ref{sec:randd}. Section~\ref{sec:summary} contains the summary.


\section{ Formalism}\label{sec:form}

The weak effective Hamiltonian $H_{{\it eff}}$ for the decays $B_c^+ \to \chi_{cJ} (P, V)^+$
is given by~\cite{Buchalla:1995vs}
\beq
H_{\it eff}\, &=&\, \frac{G_F}{\sqrt{2}}
\biggl\{ V^*_{cb}V_{uq} [ C_1(\mu) O_1(\mu)
+C_2(\mu) O_2(\mu) ] \biggr\}+ {\rm H.c.}\;,
\label{eq:heff}
\eeq
with the Fermi constant $G_F=1.16639\times 10^{-5}{\rm GeV}^{-2}$, the Cabibbo-Kobayashi-Maskawa
(CKM) matrix elements $V_{cb}$ and $V_{uq}$~\cite{Cabibbo:1963yz,Kobayashi:1973fv},
$q$ being the light quark $d$ or $s$, and the local four-quark tree operators
\beq
O_1 \, &=&\,
\bar{q}_\alpha \gamma_\mu (1 - \gamma_5) u_\beta\; \bar{c}_\beta \gamma_\mu (1 - \gamma_5) b_\alpha \;,
\qquad
O_2 \, =\, \bar{q}_\alpha \gamma_\mu (1 - \gamma_5) u_\alpha\; \bar{c}_\beta \gamma_\mu (1 - \gamma_5) b_\beta \;.
\label{eq:operators}
\eeq
The Wilson coefficients $C_i(\mu)$ sum the large logarithms $\ln (m_W/\mu)$ to all orders
in the strong coupling $\alpha_s$, where $m_W$ is the $W$ boson mass and $\mu$ is the
renormalization scale. We will not address {\it CP} violation here due to the absence of
penguin operators.

\begin{figure}[!!htb]
\centering
\begin{tabular}{l}
\includegraphics[width=0.95\textwidth]{fig1a}
\end{tabular}
\caption{Leading-order Feynman diagrams for the decays $B_c^+ \to \chi_{cJ}\ (P, V)^+$
in the iPQCD framework, with (a), (b) and (c), (d) contributing to the factorizable and
nonfactorizable emission amplitudes correspondingly. }
\label{fig:fig1}
\end{figure}

Figure~\ref{fig:fig1} depicts the leading-order Feynman diagrams for the
$B_c^+ \to \chi_{cJ} (P, V)^+$ decays in the iPQCD framework, whose
evaluation is similar to that of $B_c^+ \to J/\psi M^+$. The factorization of
the $B_c^+ \to \chi_{cJ} (P, V)^+$ decay amplitudes is formulated, according
to~\cite{Liu:2023kxr}, as
\beq
A(B_c^+ \to \chi_{cJ} (P, V)^+)
&\equiv& \langle \chi_{cJ} (P, V)^+ | H_{\it eff} | B_c^+ \rangle \non
&\sim &\int\!\! d x_1 d
x_2 d x_3 b_1 d b_1 b_2 d b_2 b_3 d b_3
\non && \cdot {\mathrm{Tr}}
\left [ C(t) \Phi_{B_c}(x_1, b_1) \Phi_{P,V}(x_2, b_2)
\Phi_{\chi_{cJ}}(x_3, b_3) H(x_i, b_i, t) e^{-S(t)} \right ]\;,
\label{eq:a2}
\eeq
where $x_i$ are the longitudinal quark momentum fractions, the variables $b_i$ are conjugate
to the quark transverse momenta $k_{iT}$, $t$ is set to the largest running energy scale
involved in the hard kernel $H$, Tr is the abbreviation of the traces over Dirac and color
indices, and the wave functions $\Phi$ describe the quark momentum distributions within
mesons. The Sudakov factor $e^{-S(t)}$, whose explicit expression was derived
in~\cite{Botts:1989kf,Liu:2018kuo,Liu:2020upy,Liu:2023kxr}, strongly suppresses the
long-distance (large-$b$) contributions~\cite{Botts:1989kf,Keum:2000ph}.
We emphasize that the present formalism has included the charm quark mass effects
into the Sudakov factor, compared to the conventional PQCD approach
to the $B_c^+ \to \chi_{cJ} (P, V)^+$ decays~\cite{Rui:2017pre}.

We specify the meson distribution amplitudes needed for implementing the factorization formula
in Eq.~(\ref{eq:a2}). The knowledge about the $\chi_{cJ}$ meson distribution amplitudes
$\Phi_{\chi_{cJ}}(x)$, whose dependence on the intrinsic $b$ dependence has been ignored,
is still quite limited. We employ the models in Refs.~\cite{Chen:2005ht,Wang:2007fs,
Rui:2017pre}. The twist-2 distribution amplitude $\phi_{\chi_{c0}}^v(x)$
and the twist-3 one $\phi_{\chi_{c0}}^s(x)$ of the scalar $\chi_{c0}$ are defined via
\beq
\Phi_{\chi_{c0}}(x) & = & \frac{i}{\sqrt{2 N_c}}
\biggl\{\psl\; \phi_{\chi_{c0}}^v(x)+ m_{\chi_{c0}} \phi_{\chi_{c0}}^s(x) \biggr\}\;,
\eeq
with the $\chi_{c0}$ meson momentum $P$ and mass $m_{\chi_{c0}}$, and the charm quark
momentum fraction $x$. They are assumed to take the asymptotic forms
\beq
\phi^v_{\chi_{c0}}(x)&=& 53.74 \frac{f_{\chi_{c0}}}{2\sqrt{2 N_c}}
\biggl[  x(1-x)(1-2x) \biggr] {\cal C}(x)\;,
\label{eq:c0-t2}
\\
\phi^s_{\chi_{c0}}(x)&=& 2.12 \frac{f_{\chi_{c0}}}{2\sqrt{2N_c}}  {\cal C}(x)\;,
\label{eq:c0-t3}
\eeq
with the decay constant $f_{\chi_{c0}}$. The function ${\cal C}(x)$ is extracted from the
$p$-wave Schr$\ddot{\rm o}$dinger states in a Coulomb potential (details can be found
in the appendix of Ref.~\cite{Rui:2017pre}),
\beq
{\cal C}(x) &=& \biggl\{[x(1-x)(1-4x(1-x))^3]^{1/2}\Big{/}[1-4x(1-x)(1-v^2/4)]^2 \biggr\}^{1-v^2}\;,
\eeq
where $v^2 =0.3$~\cite{Chen:2005ht} stands for small relativistic corrections to the
Coulomb wave functions~\cite{Bondar:2004sv}.
The distribution amplitudes $\phi_{\chi_{c0}}^v(x)$ and $\phi_{\chi_{c0}}^s(x)$ are
antisymmetric and symmetric, respectively, to satisfy the charge conjugation invariance
of scalars~\cite{Chernyak:1983ej,Diehl:2001xe}.

Note that the numerical
coefficients of the charmonium distribution amplitudes in Eqs.~(\ref{eq:c0-t2}) and (\ref{eq:c0-t3}) are chosen to
meet the normalization conditions,
\beq
\int_0^1 dx \phi_{\rm sym}(x) &=& \frac{f}{2\sqrt{2 N_c}}\;,
\eeq
for the symmetric distribution amplitudes, and
\beq
\int_0^1 dx \phi_{\rm asym}(x) (1-2x) &=& \frac{f}{2\sqrt{2 N_c}}\;,
\eeq
for the anti-symmetric ones, where $f$ denotes a decay constant. Such numerical factors
appear, because the charmonium distribution amplitudes are not expanded in terms of
the Gegenbauer polynomials like light meson distribution amplitudes.

For the axial-vector $\chi_{c1}$, the twist-2 distribution amplitude
$\phi_{\chi_{c1}}^{L(T)}(x)$ and the twist-3 distribution amplitude
$\phi_{\chi_{c1}}^{t(v)}(x)$ associated with the longitudinal (transverse) polarization
are defined via
\beq
\Phi_{\chi_{c1}}^L(x) & = & \frac{1}{\sqrt{2 N_c}} \gamma_5
\biggl\{m_{\chi_{c1}}\, {\epsl}_{\chi_{c1}}^L \,\phi_{\chi_{c1}}^L (x)
+ {\epsl}_{\chi_{c1}}^L \, \psl\; \phi_{\chi_{c1}}^t(x) \biggr\}\;,
 \\
\Phi^{T}_{\chi_{c1}}(x) &=&  \frac{1}{\sqrt{2 N_c}} \gamma_5
\biggl\{ m_{\chi_{c1}}\, {\epsl}_{\chi_{c1}}^T\, \phi_{\chi_{c1}}^v(x)
+ {\epsl}_{\chi_{c1}}^T\, \psl\; \phi_{\chi_{c1}}^T(x) \biggr\}\;,
\eeq
where $m_{\chi_{c1}}$ is the $\chi_{c1}$ meson mass, and $\epsilon^{L}$ ($\epsilon^{T}$)
denotes the longitudinal (transverse) polarization vector of $\chi_{c1}$.
These distribution amplitudes are written as
\beq
\phi_{\chi_{c1}}^L(x) &=& 12.60 \frac{f_{\chi_{c1}}}{2\sqrt{2N_c}} \biggl[ x(1-x) \biggr] {\cal C}(x)\;,
\label{eq:c1-t2-L}
\\
\phi_{\chi_{c1}}^T(x) &=& 53.74 \frac{f_{\chi_{c1}}^{\perp}}{2\sqrt{2N_c}}
\biggl[ x(1-x)(1-2x) \biggr] {\cal C}(x) \;,
\label{eq:c1-t2-T}
\\
\phi_{\chi_{c1}}^t(x) &=& 23.16 \frac{f_{\chi_{c1}}^{\perp}}{2\sqrt{2N_c}}
\biggl[ (1-2x) (1-6x+6x^2) \biggr]{\cal C}(x)\;,
\label{eq:c1-t3-t}
\\
\phi_{\chi_{c1}}^v(x) &=& 1.59 \frac{f_{\chi_{c1}}}{2\sqrt{2N_c}} \biggl[1+(1-2x)^2 \biggr] {\cal C}(x)\;,
\label{eq:c1-t3-v}
\eeq
with the decay constants $f_{\chi_{c1}}$ and $f_{\chi_{c1}}^\perp$.

A similar decomposition applies to the distribution amplitudes for longitudinally and
transversely polarized tensors $\chi_{c2}$,
\beq
\Phi^{L}_{\chi_{c2}}(x) &=&  \frac{1}{\sqrt{2 N_c}}
\biggl\{ m_{\chi_{c2}}\, {\epsl}_{\chi_{c2}}^L \,\phi_{\chi_{c2}}^L(x)
+ {\epsl}_{\chi_{c2}}^L \, \psl\; \phi_{\chi_{c2}}^t(x) \biggr\}\;,
 \\
\Phi^{T}_{\chi_{c2}}(x) &=&  \frac{1}{\sqrt{2 N_c}}
\biggl\{ m_{\chi_{c2}}\, {\epsl}_{\chi_{c2}}^T\, \phi_{\chi_{c2}}^v(x)
+ {\epsl}_{\chi_{c2}}^T\, \psl\; \phi_{\chi_{c2}}^T(x) \biggr\}\;,
 \eeq
respectively, with the twist-2 components
\beq
\phi_{\chi_{c2}}^L(x) &=& 53.74 \frac{f_{\chi_{c2}}}{2\sqrt{2N_c}}
\biggl[ x(1-x)(1-2x) \biggr] {\cal C}(x) \;,
\label{eq:c2-t2-L}
\\
\phi_{\chi_{c2}}^T(x) &=& 53.74 \frac{f_{\chi_{c2}}^\perp}{2\sqrt{2N_c}}
\biggl[ x(1-x)(1-2x) \biggr] {\cal C}(x) \;,
\label{eq:c2-t2-T}
\eeq
and the twist-3 ones
\beq
\phi_{\chi_{c2}}^t(x) &=& 23.16 \frac{f_{\chi_{c2}}^{\perp}}{2\sqrt{2N_c}}
\biggl[ (1-2x)(1-6x+6x^2) \biggr] {\cal C}(x)\;,
\label{eq:c2-t3-t}
\\
\phi_{\chi_{c2}}^v(x) &=& 12.44 \frac{f_{\chi_{c2}}}{2\sqrt{2N_c}}
\biggl[(1-2x)^3 \biggr]  {\cal C}(x)\;,
\label{eq:c2-t3-v}
\eeq
$m_{\chi_{c2}}$ being the $\chi_{c2}$ meson mass, and $f_{\chi_{c2}}$ and $f_{\chi_{c2}}^\perp$
being the decay constants.


The $B_c$ meson wave function $\Phi_{B_c}(x, b)$ and the distribution amplitudes
$\Phi_{P,V}(x)$ of the light pseudoscalar and vector mesons have been collected
in~\cite{Liu:2023kxr}, which are not shown for simplicity.
For the inputs, such as the decay constants, the shape parameter $\beta_{B_c}$
in $\Phi_{B_c}(x, b)$ and the Gegenbauer moments in $\Phi_{P,V}(x)$ refer
to~\cite{Liu:2023kxr}. The related measurements, as well as lattice QCD or QCD
sum rule calculations, are expected to enhance the accuracy of the above nonperturbative
quantities.

We construct the $B_c^+ \to \chi_{cJ} (P, V)^+$ decay amplitudes in Eq.~(\ref{eq:a2})
by convoluting the hard kernels $H(x_i,b_i,t)$ from Fig.~\ref{fig:fig1} with the
aforementioned meson distribution amplitudes. The factorization formulas read
\beq
A &=& V_{cb}^* V_{uq} \biggl[(\frac{1}{3} C_1 + C_2) F_e \cdot f_{P} + C_1 M_e \biggr] 
\label{eq:DecAmp-P}
\eeq
for $B_c^+ \to \chi_{cJ} P^+$ and $B_c^+ \to \chi_{c0} V^+$, and
\beq
A^h &=& V_{cb}^* V_{uq} \biggl[(\frac{1}{3} C_1 + C_2) F_e^{h} \cdot f_{V} + C_1  M_e^{h} \biggr] 
\label{eq:DecAmp-V}
\eeq
for $B_c^+ \to \chi_{c1,c2} V^+$ with the vector meson polarizations $h = L, N, T$, where
$f_{P(V)}$ represents the decay constant of the light pseudoscalar (vector) meson. The
expressions of the factorizable (nonfactorizable) emission amplitudes $F_{e}^{(h)}$
($M_{e}^{(h)}$) induced by the $(V-A)(V-A)$ operators in Eq.~(\ref{eq:operators}) have been
provided in Eqs.~(32)$-$(48) of~\cite{Rui:2017pre}. The corresponding BR is then given, in
the rest frame of a heavy $B_c$ meson by
\beq
{\rm BR}(B_c^+ \to \chi_{cJ} (P, V)^+)&\equiv&
\tau_{B_c}\cdot \Gamma(B_c^+ \to \chi_{cJ} (P, V)^+)
\non
&=& \tau_{B_c}\cdot\frac{G_{F}^{2}|\bf{P_c}|}{16 \pi m^{2}_{B_c} }
|A(B_c^+ \to \chi_{cJ} (P, V)^+)|^2\;,
\label{eq:BR-def}
\eeq
where $\tau_{B_c} = 0.507$ ps is the $B_c$ meson lifetime, $\Gamma$ is the decay width, and
$|{\bf P}_c|\equiv |{\bf P}_{P(V)}|=|{\bf P}_{\chi_{cJ}}|
= \sqrt{\lambda(m_{B_c}^2, m_{P (V)}^2, m_{\chi_{cJ}}^2)}/(2 m_{B_c})$ is the
momentum of either the $P (V)$ or $\chi_{cJ}$ meson in the final state,
with the $B_c$ ($P (V)$) meson mass $m_{B_c}$ ($m_{P(V)}$)
and the ${\rm K\ddot{a}ll\acute{e}n}$ function
$\lambda(x, y, z) = x^2 +y^2 +z^2 - 2 xy - 2 xz - 2 yz$~\cite{ParticleDataGroup:2024cfk}.

\section{Numerical Results and Discussions} \label{sec:randd}

To proceed with the numerical study, we adopt the following QCD scale and meson masses
(in units of GeV)~\cite{Keum:2000ph,ParticleDataGroup:2024cfk},
\beq
 \Lambda_{\overline{\mathrm{MS}}}^{(f=4)} &=& 0.250\; ,
 \qquad m_W = 80.41\;,
 \qquad m_{B_c} = 6.275 \;,
 \qquad m_{\chi_{c2}} = 3.556\;,
 \non
m_{\chi_{c1}} &=& 3.511\;,
\qquad m_{\chi_{c0}} = 3.415\;,
 \qquad  m_b = 4.8 \;,
 \qquad  m_{c}= 1.5 \;,
\label{eq:mass}
\eeq
and the CKM matrix elements~\cite{ParticleDataGroup:2024cfk}
\beq
|V_{cb}| &=& 0.04182^{+0.00085}_{-0.00074}\;,
\quad
|V_{ud}| = 0.97435 \pm 0.00016\;,
\quad
|V_{us}| = 0.22500 \pm 0.00067\;.
\eeq
The $\chi_{cJ}$ meson decay constants at the scale
$\mu =m_c$ are set to (in units of GeV)~\cite{Olpak:2016wkf}
\beq
f_{\chi_{c0}} &=& 0.093 \pm 0.009\;,
\qquad
f_{\chi_{c1}} = 0.185 \pm 0.019\;,
\qquad
f_{\chi_{c1}}^\perp = 0.090 \pm 0.009\;,
\non
f_{\chi_{c2}} &=&  0.181 \pm 0.018\;,
\qquad
f_{\chi_{c2}}^\perp = 0.131 \pm 0.013\;,
\label{eq:DCs-chi}
\eeq
for which $10\%$ variations of the central values have been assigned for assessing the
potential theoretical errors. As stated in \cite{Liu:2018kuo}, the charmonium distribution
amplitudes are defined at the charm quark mass scale in the iPQCD framework, so are the
associated decay constants. Note that the decay constants in Ref.~\cite{Olpak:2016wkf}
were derived at the scale $m_c = 1.628$ GeV, while the charm quark mass is chosen as
$m_c=1.5$ GeV in Eq.~(\ref{eq:mass}). Therefore, the renormalization-group
evolution~\cite{Braguta:2008qe}
\beq
f_{c\bar c}(\mu) &=& f_{c\bar c}(\mu_0)
\Biggl(\frac{\alpha_s(\mu)}{\alpha_s(\mu_0)}\Biggr)^{\gamma/b_0}\;,
\eeq
with the coefficient $b_0 = 11 - 2n_f/3$, $n_f$ being the active flavor number,
should be taken into account. The anomalous dimensions take
$\gamma = 8C_F/3$ for $f_{\chi_{c0}}$ and $f_{\chi_{c2}}$ with the color factor
$C_F=4/3$, $\gamma = 3 C_F$ for  $f^\perp_{\chi_{c1}}$ and $f^\perp_{\chi_{c2}}$,
and $\gamma =0$ for $f_{\chi_{c1}}$~\cite{Braguta:2008qe}.
Equation~(\ref{eq:DCs-chi}) represents the results of the evolution from 1.628 GeV down
to 1.5 GeV.

\subsection{\boldmath  $ B_c^+ \to \chi_{cJ} P^+$}
\label{ssec:chicjP}

We present the iPQCD predictions for the BRs and ratios associated with the decays
$B_c^+ \to \chi_{cJ} (P, V)^+$. The {\it CP}-averaged $B_c^+ \to \chi_{c2} P^+$ BRs are
written as
\beq
{\rm BR}(B_c^+ \to \chi_{c2} \pi^+)&=&
3.75
^{+1.07}_{-0.80}(\beta_{B_c})
^{+0.91}_{-0.80}(f_{c\bar c})
^{+0.01}_{-0.00}(a_\pi)
^{+0.16}_{-0.13}(V_{cb})
\times 10^{-4} \;,
\label{eq:c2pi}
\\
{\rm BR}(B_c^+ \to \chi_{c2} K^+) &=&
2.75
^{+0.74}_{-0.57}(\beta_{B_c})
^{+0.66}_{-0.59}(f_{c\bar c})
^{+0.22}_{-0.21}(a_K)
^{+0.11}_{-0.10}(V_{cb})
\times 10^{-5}\;,
\label{eq:c2k}
\eeq
where the errors mainly arise from the shape parameter $\beta_{B_c}$ and the less
constrained decay constants $f_{\chi_{c2}}$ and $f_{\chi_{c2}}^{\perp}$.
As mentioned in the Introduction, the decays $B_c^+ \to \chi_{c2} P^+$
have been investigated extensively in the literature~\cite{Chang:2001pm,
Kiselev:2001zb,LopezCastro:2002woj,Ivanov:2006ni,Hernandez:2006gt,
Ebert:2010zu,Wang:2011jt,Rui:2017pre,Zhu:2017lwi,Chen:2021vmb,Losacco:2023uvp,Lu:2025bvi}
with dramatically different outcomes. Our prediction
for ${\rm BR}(B_c^+ \to \chi_{c2} \pi^+)$ is consistent with those from
Refs.~\cite{Ivanov:2006ni,Ebert:2010zu} in Table~\ref{tab:Rs-c2}
within uncertainties. The value in Eq.~(\ref{eq:c2pi}) is lower than those
in the conventional PQCD approach by an order of magnitude~\cite{Rui:2017pre}
and in the leading-order NRQCD factorization by a factor of about $25$~\cite{Chen:2021vmb}.
Owing to the unclear production cross section for $B_c$ mesons, one has to consider
relative ratios between BRs of $B_c$ meson decays. As shown in Eq.~(\ref{eq:r-c2oj-e}),
the $B_c^+ \to \chi_{c2} \pi^+$ channel was observed in terms of its BR
relative to the $B_c^+ \to J/\psi \pi^+$ one. Combined with the
$B_c^+ \to J/\psi \pi^+$ BR in the iPQCD formalism~\cite{Liu:2023kxr},
the ratio is given by
\beq
R_{\chi_{c2}/{J/\psi}}
=  0.32 \pm 0.05\;,
\label{eq:r-c2oj-t}
\eeq
where errors from various sources have been added in quadrature. It is compatible with
the data in Eq.~(\ref{eq:r-c2oj-e}), and also with the
predictions in~\cite{Ivanov:2006ni,Hernandez:2006gt}.

We obtain the $B_c^+ \to \chi_{c0} P^+$ and $B_c^+ \to\chi_{c1} P^+$ BRs,
\beq
{\rm BR}(B_c^+ \to \chi_{c0} \pi^+)&=&
2.02
^{+0.57}_{-0.41}(\beta_{B_c})
^{+0.41 }_{-0.37 }(f_{c\bar c})
^{+0.00}_{-0.00}(a_\pi)
^{+0.08}_{-0.07}(V_{ub})
\times 10^{-4} \;,
\label{eq:c0pi}
\\
{\rm BR}(B_c^+ \to \chi_{c0} K^+) &=&
1.47
^{+0.40}_{-0.29}(\beta_{B_c})
^{+0.30 }_{-0.27 }(f_{c\bar c})
^{+0.12}_{-0.11}(a_K)
^{+0.06}_{-0.05}(V_{cb})
\times 10^{-5}\;,
\eeq
and
\beq
{\rm BR}(B_c^+ \to \chi_{c1} \pi^+)&=&
4.16
^{+0.68}_{-0.62}(\beta_{B_c})
^{+1.26}_{-1.12}(f_{c\bar c})
^{+0.00}_{-0.02}(a_\pi)
^{+0.17}_{-0.14}(V_{cb})
\times 10^{-5} \;,
\label{eq:c1pi}
\\
{\rm BR}(B_c^+ \to \chi_{c1} K^+) &=&
2.65
^{+0.37}_{-0.36}(\beta_{B_c})
^{+0.83}_{-0.74}(f_{c\bar c})
^{+0.67}_{-0.59}(a_K)
^{+0.10}_{-0.10}(V_{cb})
\times 10^{-6}\;,
\eeq
respectively. It is seen from Table~\ref{tab:Rs-c0} that the iPQCD result for
${\rm BR}(B_c^+ \to \chi_{c0} \pi^+)$ matches those in Refs.~\cite{Hernandez:2006gt,Ebert:2010zu}
and in \cite{Lu:2025bvi} without Coulomb-like corrections.
The decay $B_c^+ \to \chi_{c1} \pi^+$ has not been observed, and the upper bound of
its BR has been imposed by the LHCb as in Eq.~(\ref{eq:r-c1oc2-e}).
Our predicted ${\rm BR}(B_c^+ \to \chi_{c1} \pi^+)$ agrees with
those from Refs.~\cite{Zhu:2017lwi} and~\cite{Lu:2025bvi} with Coulomb-like corrections in Table~\ref{tab:Rs-c1}.
We also compute the ratio between the $B_c^+ \to \chi_{c1} \pi^+$ and
$B_c^+ \to \chi_{c2} \pi^+$ BRs,
\beq
R_{\chi_{c1}/\chi_{c2}}
= 0.11^{+0.01}_{-0.01} \;,
\label{eq:r-c1oc2-p}
\eeq
which satisfies Eq.~(\ref{eq:r-c1oc2-e}). It has been pointed out that most of the
predictions for this ratio, spanning a wide range as listed in Table~\ref{tab:Rs-c1},
respect Eq.~(\ref{eq:r-c1oc2-e}). More precise data will help discriminate the different
theoretical approaches.

The evidence of the $B_c^+ \to \chi_{c0} \pi^+$ mode was reported in association with the
ratio $\sigma(B_c^+)/\sigma(B^+)$ of the $B_c^+$ and $B^+$ production cross sections
in Eq.~(\ref{eq:BR-c0pi}). Note that these cross sections include contributions
from excited states. The same ratio $\sigma(B_c^+)/\sigma(B^+)$ appears in the
measurements by the LHCb~\cite{LHCb:2012ihf,LHCb:2014mvo},
\beq
R_{c/u}^{\rm LHCb} &\equiv& \frac{\sigma(B_c^+)}{\sigma(B^+)} \cdot
\frac{{\rm BR}(B_c^+ \to J/\psi \pi^+)}{{\rm BR}(B^+ \to J/\psi K^+)} =
\left\{ \begin{array}{ll}
(0.61 \pm 0.12) \%  \qquad (p_T > 4~{\rm GeV}, 2.5 < \eta < 4.5)
& \vspace{0.12cm}
\\
(0.698 \pm 0.023) \%  \quad (4 <p_T < 20~{\rm GeV}, 2.5 < \eta < 4.5)
& \vspace{0.12cm}
\\
(0.683 \pm 0.020) \%  \quad (0 <p_T < 20~{\rm GeV}, 2.0 < y < 4.5)
&  \\ \end{array} \right.,
\label{eq:R-cou-LHCb}
\eeq
and by the CMS and ATLAS collaborations~\cite{CMS:2014oqy,ATLAS:2019jpi},
\beq
R_{c/u}^{\rm CMS} &\equiv& \frac{\sigma(B_c^+)}{\sigma(B^+)} \cdot
\frac{{\rm BR}(B_c^+ \to J/\psi \pi^+)}{{\rm BR}(B^+ \to J/\psi K^+)}
= (0.48 \pm 0.08) \%  \quad (p_T > 15~{\rm GeV}, |y| < 1.6)\;,
\label{eq:R-cou-CMS}
\\
R_{c/u}^{\rm ATLAS} &\equiv& \frac{\sigma(B_c^+)}{\sigma(B^+)} \cdot
\frac{{\rm BR}(B_c^+ \to J/\psi \pi^+)}{{\rm BR}(B^+ \to J/\psi K^+)}
= (0.34^{+0.07}_{-0.05}) \%  \qquad ( p_T > 13~{\rm GeV}, |y| < 2.3)\;,
\label{eq:R-cou-ATLAS}
\eeq
with the bottom-meson transverse momentum $p_T$ and (pseudo)rapidity $(\eta) y$.

Given the measured ${\rm BR}(B^+ \to J/\psi K^+)=
(1.020 \pm 0.019) \times 10^{-3}$~\cite{ParticleDataGroup:2024cfk}, we extract
$\frac{\sigma(B_c^+)}{\sigma(B^+)}\times {\rm BR}(B_c^+ \to J/\psi \pi^+)
= (6.97 \pm 0.13) \times 10^{-6}$ from Eq.~(\ref{eq:R-cou-LHCb}) in the fiducial region
corresponding to the transverse momentum $0 <p_T < 20~{\rm GeV}$ and the rapidity
$2.0 < y < 4.5$. The LHCb data $\frac{\sigma(B_c^+)}{\sigma(B^+)}
\times {\rm BR}(B_c^+ \to \chi_{c0}\pi^+) = (9.8^{+3.5}_{-3.1}) \times 10^{-6}$
from the same fiducial region in Eq.~(\ref{eq:BR-c0pi}) then yields~\footnote{
To derive the error $\delta_X$ associated with a quantity $X= u/v$, we follow the
formula $\delta_X^2 = (u^2\delta_v^2+v^2\delta_u^2)/v^4$, where $\delta_u$ ($\delta_v$)
represents the uncertainty of $u$ ($v$).}
\beq
R^{\rm Exp}_{\chi_{c0}/J/\psi} &\equiv&
\frac{{\rm BR}(B_c^+ \to \chi_{c0} \pi^+)}{{\rm BR}(B_c^+ \to J/\psi \pi^+)}
= \frac{\frac{\sigma(B_c^+)}{\sigma(B^+)} \times
{\rm BR}(B_c^+ \to \chi_{c0} \pi^+)}{\frac{\sigma(B_c^+)}{\sigma(B^+)}
\times {\rm BR}(B_c^+ \to J/\psi \pi^+)}
=  1.41^{+0.50}_{-0.45}\;.
\label{eq:R-c0oj-e}
\eeq
Our prediction for the above ratio
\beq
R_{\chi_{c0}/J/\psi}
=  0.17^{+0.02}_{-0.02} \;,
\label{eq:r-c0oj-t}
\eeq
seems to deviate from Eq.~(\ref{eq:R-c0oj-e}). On one hand, the data suffer large
uncertainties. On the other hand, the nonperturbative inputs related to a $\chi_{c0}$
meson are still poorly known. If the decay constant $f_{\chi_{c0}}$ takes a value around
0.3 GeV~\cite{Novikov:1977dq,Chen:2005ht,Lu:2025bvi}, the $B_c^+ \to \chi_{c0} \pi^+$ BR in
Eq.~(\ref{eq:r-c0oj-t}) can be greatly amplified by a factor of 10,
and the consistency with the data will be improved.

Analogous to Eqs.~(\ref{eq:r-c1oc2-p}) and (\ref{eq:r-c0oj-t}), we consider the ratios,
\beq
R_{\chi_{c1}/J/\psi} &\equiv&
\frac{{\rm BR}(B_c^+ \to \chi_{c1} \pi^+)}{{\rm BR}(B_c^+ \to J/\psi \pi^+)}
= (3.56^{+0.81}_{-0.82}) \times 10^{-2}\;,
\label{eq:r-c1oj-t}
\eeq
which agrees with the predictions from Refs.~\cite{Chang:2001pm,Ivanov:2006ni,Lu:2025bvi}
in Table~\ref{tab:Rs-c1}, and
\beq
R_{\chi_{c0}/\chi_{c2}} &\equiv&
\frac{{\rm BR}(B_c^+ \to \chi_{c0} \pi^+)}{{\rm BR}(B_c^+ \to \chi_{c2} \pi^+)}
= 0.54^{+0.02}_{-0.02} \;,
\label{eq:r-c0oc2-p}
\\
R_{\chi_{c1}/\chi_{c0}} &\equiv&
\frac{{\rm BR}(B_c^+ \to \chi_{c1} \pi^+)}{{\rm BR}(B_c^+ \to \chi_{c0} \pi^+)}
= 0.21^{+0.01}_{-0.04} \;,
\label{eq:r-c1oc0-p}
\eeq
which await future experimental scrutiny.

The ratio between the $B_c^+ \to (c\bar c) K^+$ and $B_c^+ \to (c\bar c) \pi^+$ BRs,
governed by the factorizable emission amplitudes~\cite{Liu:2023kxr}, is simplified into
\beq
R_{K/\pi} &\equiv&
\frac{{\rm BR}(B_c^+ \to (c\bar c) K^+)}{{\rm BR}(B_c^+ \to (c\bar c) \pi^+)}
\approx
\frac{f_K^2}{f_\pi^2}\cdot \frac{|V_{us}|^2}{|V_{ud}|^2}
 = (7.95^{+0.05}_{-0.01}) \times 10^{-2}\;,
\label{eq:rpkcc-n}
\eeq
with the kaon (pion) decay constant $f_K$ ($f_\pi$). The observed departures from
Eq.~(\ref{eq:rpkcc-n}) signify the nonfactorizable contributions in the considered
channels. To be specific, we present the ratios between the $B_c^+ \to \chi_{cJ} K^+$
and $B_c^+ \to \chi_{cJ} \pi^+$ BRs,
\beq
R_{K/\pi}^{\chi_{c2}}&\equiv& \frac{{\rm BR}(B_c^+ \to \chi_{c2} K^+)}
{{\rm BR}(B_c^+ \to \chi_{c2} \pi^+)}=
(7.33^{+0.57}_{-0.57})
\times 10^{-2}
\;,
\label{eq:rkpc2-t}
\\
R_{K/\pi}^{\chi_{c0}} &\equiv&
\frac{{\rm BR}(B_c^+ \to \chi_{c0} K^+)}{{\rm BR}(B_c^+ \to \chi_{c0} \pi^+)}
=(7.28^{+0.68}_{-0.56}) \times 10^{-2}
\;,
\label{eq:rkpc0-t}
\\
R_{K/\pi}^{\chi_{c1}} &\equiv&
\frac{{\rm BR}(B_c^+ \to \chi_{c1} K^+)}{{\rm BR}(B_c^+ \to \chi_{c1} \pi^+)}
=(6.37^{+1.61}_{-1.40}) \times 10^{-2}
\;,
\label{eq:rkpc1-t}
\eeq
where the uncertainties are dominated by the variation of the Gegenbauer moment $a_1^K$
in the twist-2 kaon distribution amplitude. Though Eqs.~(\ref{eq:rkpc2-t})$-$(\ref{eq:rkpc1-t})
basically align with the naive expectation in Eq.~(\ref{eq:rpkcc-n}), they reveal a
dependence of the nonfactorizable emission contributions on the $\chi_{cJ}$ final states.

The BESIII Collaboration reported their latest measured BRs of the strong decays
$\chi_{c0,c2} \to \pi^+ \pi^- $ and $K^+ K^-$~\cite{BESIII:2025tho},
\beq
{\rm BR}(\chi_{c0} \to \pi^+ \pi^-) &=& (0.606 \pm 0.015)\%  \;, \qquad
{\rm BR}(\chi_{c0} \to K^+ K^-) = (0.636 \pm 0.015)\%\;,
\label{eq:c0-ppkk}
\\
{\rm BR}(\chi_{c2} \to \pi^+ \pi^-) &=& (0.161 \pm 0.005)\% \;, \qquad
{\rm BR}(\chi_{c2} \to K^+ K^-) = (0.122 \pm 0.004)\%\;,
\label{eq:c2-ppkk}
\eeq
where errors from various sources have been added in quadrature. With these data, one
can deduce BRs of multibody $B_c$ meson decays via charmonium resonances under
the narrow-width approximation. The $B_c^+ \to \pi^+ \chi_{c0,c2} (\to \pi^+ \pi^- / K^+ K^-)$
BRs are then found to be
\beq
{\rm BR}(B_c^+ \to \chi_{c0} (\to \pi^+ \pi^-) \pi^+) &\equiv&
{\rm BR}(B_c^+ \to \chi_{c0} \pi^+)\cdot {\rm BR}(\chi_{c0} \to \pi^+ \pi^-)
=  (1.22^{+0.43}_{-0.34} ) \times 10^{-6}, \\
{\rm BR}(B_c^+ \to \chi_{c0} (\to K^+ K^-) \pi^+) &\equiv&
{\rm BR}(B_c^+ \to \chi_{c0} \pi^+)\cdot {\rm BR}(\chi_{c0} \to K^+ K^-)
= (1.28^{+0.45}_{-0.36}) \times 10^{-6},
\label{eq:br-c0-kk}
\\
{\rm BR}(B_c^+ \to \chi_{c2} (\to \pi^+ \pi^-) \pi^+) &\equiv&
{\rm BR}(B_c^+ \to \chi_{c2} \pi^+)\cdot {\rm BR}(\chi_{c2} \to \pi^+ \pi^-)
= (0.60^{+0.23}_{-0.18}) \times 10^{-6},
\\
{\rm BR}(B_c^+ \to \chi_{c2} (\to K^+ K^-) \pi^+) &\equiv&
{\rm BR}(B_c^+ \to \chi_{c2} \pi^+)\cdot {\rm BR}(\chi_{c2} \to K^+ K^-)
= (0.46^{+0.17}_{-0.14}) \times 10^{-6}.
\eeq
The above results suggest that the $B_c^+ \to \chi_{c0,c2} (\to \pi^+ \pi^-) \pi^+$
and $B_c^+ \to \chi_{c0} (\to K^+ K^-) \pi^+$ decays with larger BRs
are observable in the near future. Employing the data in Eq.~(\ref{eq:BR-c0pi})
and the predictions in Eqs.~(\ref{eq:c0pi}, \ref{eq:br-c0-kk}), we attain the ratio
\beq
\frac{\sigma(B_c^+)}{\sigma(B^+)}\cdot {\rm BR}(B_c^+ \to \chi_{c0} (\to K^+ K^-) \pi^+)
= (4.98^{+3.05}_{-2.47}) \times 10^{-8}\;,
\eeq
in line with the measured one $(8.0^{+4.4}_{-3.8}) \times 10^{-8}$ within large
uncertainties~\cite{LHCb:2016utz}. More precise predictions and data are required for a
serious comparison.

The BESIII Collaboration also reported the improved measurements of the decays
$\chi_{cJ} \to 2(\pi^+ \pi^-)$~\cite{BESIII:2024jmr}. The
$\chi_{cJ} \to \pi^+ \pi^- K^+ K^-$ BRs are available in Ref.~\cite{ParticleDataGroup:2024cfk}.
The inputs of these four-body strong decays,
\beq
{\rm BR}(\chi_{c0} \to 2(\pi^+\pi^-)) &=& (2.127 \pm 0.101)\% \;,
\quad
{\rm BR}(\chi_{c0} \to \pi^+\pi^- K^+K^-) = (1.82 \pm 0.16)\% \;,
\label{eq:c0-2ppkk}
\\
{\rm BR}(\chi_{c1} \to 2(\pi^+\pi^-)) &=& (0.685 \pm 0.031)\% \;,
\quad
{\rm BR}(\chi_{c1} \to \pi^+\pi^- K^+K^-) = (0.45 \pm 0.10) \%\;,
\label{eq:c1-2ppkk}
\\
{\rm BR}(\chi_{c2} \to 2(\pi^+\pi^-)) &=& (1.153 \pm 0.063)\% \;,
\quad
{\rm BR}(\chi_{c2} \to \pi^+\pi^- K^+K^-) = (0.83 \pm 0.11) \%\;,
\label{eq:c2-2ppkk}
\eeq
lead to the BRs of the five-body decays
$B_c^+ \to \pi^+ \chi_{cJ} [\to \pi^+ \pi^-(\pi^+ \pi^- / K^+ K^-)]$,
\beq
{\rm BR}(B_c^+ \to \chi_{c0} ( \to \pi^+ \pi^-\pi^+\pi^-) \pi^+) &\equiv&
{\rm BR}(B_c^+ \to \chi_{c0} \pi^+) \cdot {\rm BR}(\chi_{c0} \to 2(\pi^+ \pi^-)) \non
&=& (4.30^{+1.52}_{-1.21}) \times 10^{-6}\;,
\label{eq:br-c0-5pi}
\\
{\rm BR}(B_c^+ \to \chi_{c0} ( \to \pi^+\pi^-K^+K^-) \pi^+) &\equiv&
{\rm BR}(B_c^+ \to \chi_{c0} \pi^+) \cdot {\rm BR}(\chi_{c0} \to \pi^+ \pi^- K^+K^-)\non
&=& (3.68^{+1.33}_{-1.07}) \times 10^{-6}\;,
\label{eq:br-c0-2k3pi}
\\
{\rm BR}(B_c^+ \to \chi_{c1} ( \to \pi^+ \pi^- \pi^+\pi^-) \pi^+) &\equiv&
{\rm BR}(B_c^+ \to \chi_{c1} \pi^+) \cdot {\rm BR}(\chi_{c1} \to 2(\pi^+ \pi^-))
\non
&=& (2.85^{+0.99}_{-0.89}) \times 10^{-7}\;,
\label{eq:br-c1-5pi}
\\
{\rm BR}(B_c^+ \to \chi_{c1} ( \to \pi^+\pi^-K^+K^-) \pi^+) &\equiv&
{\rm BR}(B_c^+ \to \chi_{c1} \pi^+) \cdot {\rm BR}(\chi_{c1} \to \pi^+ \pi^- K^+K^-)
\non
&=& (1.87^{+0.77}_{-0.71}) \times 10^{-7}\;,
\label{eq:br-c1-2k3pi}
\\
{\rm BR}(B_c^+ \to \chi_{c2} (\to \pi^+ \pi^-\pi^+\pi^-) \pi^+) &\equiv&
{\rm BR}(B_c^+ \to \chi_{c2} \pi^+) \cdot {\rm BR}(\chi_{c2} \to 2(\pi^+ \pi^-))
\non
&=& (4.32^{+1.64}_{-1.34}) \times 10^{-6}\;,
\label{eq:br-c2-5pi}
\\
{\rm BR}(B_c^+ \to \chi_{c2} ( \to \pi^+\pi^- K^+K^-) \pi^+) &\equiv&
{\rm BR}(B_c^+ \to \chi_{c2} \pi^+) \cdot {\rm BR}(\chi_{c2} \to \pi^+ \pi^- K^+K^-)
\non
&=& (3.11^{+1.24}_{-1.03}) \times 10^{-6}\;,
\label{eq:br-c2-2k3pi}
\eeq
among which those of ${\cal O}(10^{-6})$ could be probed promisingly.

\subsection{\boldmath  $ B_c^+ \to \chi_{cJ} V^+$}
\label{ssec:chicjv}

We then come to the decays $B_c^+ \to \chi_{cJ} V^+$.
The $B_c^+ \to \chi_{c0} V^+$ BRs are predicted, in the iPQCD framework, as
\beq
{\rm BR}(B_c^+ \to \chi_{c0} \rho^+)&=&
2.97
^{+0.83}_{-0.59}(\beta_{B_c})
^{+0.75}_{-0.63}(f_{c\bar c})
^{+0.12}_{-0.11}(a_\rho)
^{+0.12}_{-0.10}(V_{cb})\times 10^{-4} \;,
\label{eq:c0ro}
\\
{\rm BR}(B_c^+ \to \chi_{c0} K^{*+}) &=&
2.00
^{+0.56}_{-0.40}(\beta_{B_c})
^{+0.47}_{-0.42}(f_{c\bar c})
^{+0.12}_{-0.10}(a_{K^*})
^{+0.09}_{-0.07}(V_{cb})\times 10^{-5}\;,
\label{eq:c0ks}
\eeq
to which only longitudinally polarized $V^+$ meson distribution amplitudes contribute.
It explains why ${\rm BR}(B_c^+ \to \chi_{c0} V^+)$ are in the same order as
${\rm BR}(B_c^+ \to \chi_{c0} P^+)$. Our prediction for
${\rm BR}(B_c^+ \to \chi_{c0} \rho^+)$ is close to that from Ref.~\cite{Ebert:2010zu}
in Table~\ref{tab:BRs-chicjV}. The predicted ${\rm BR}(B_c^+ \to \chi_{c0} K^{*+})$
is close to those from Refs.~\cite{Hernandez:2006gt} and \cite{Lu:2025bvi} without
the Coulomb-like corrections. The values in Eqs.~(\ref{eq:c0ro}) and (\ref{eq:c0ks})
are smaller than those in Refs.~\cite{Kiselev:2001zb,Ivanov:2006ni,Rui:2017pre,Zhang:2023ypl}
by one or even two orders of magnitude. Future measurements of
these two modes will discriminate the different theoretical methods.
We present the ratios
\beq
R^{\chi_{c0}\rho}_{J/\psi \pi} &\equiv&
\frac{{\rm BR}(B_c^+ \to \chi_{c0}\rho^+)}{{\rm BR}(B_c^+ \to J/\psi \pi^+)}
= 0.25^{+0.05}_{-0.04}
\;,
\label{eq:r-c0rojpi}
\\
R_{K^*/\rho}^{\chi_{c0}} &\equiv&
\frac{{\rm BR}(B_c^+ \to \chi_{c0} K^{*+})}{{\rm BR}(B_c^+ \to \chi_{c0} \rho^+)}
= (6.73^{+0.12}_{-0.18}) \times 10^{-2}
\;.
\label{eq:r-rhoks-c0}
\eeq
The latter, which deviates from the naive expectation $R_{K^*/\rho} \approx
|V_{us}|^2/|V_{ud}|^2 \cdot f^2_{K^*}/f_{\rho}^2 = (5.75^{+0.15}_{-0.16})\times 10^{-2}$
relying only on the factorizable emission contributions, implies the
relevance of the nonfactorizable contributions to these decays.

\begin{table}[hbt]
\caption{$B_c^+ \to \chi_{cJ} V^+$ BRs predicted in the literature.}
\label{tab:BRs-chicjV}
\begin{center}\vspace{-0.5cm}{\tiny
\begin{tabular}[t]{c|c|c|c|c|c|c|c|c|c}
\hline\hline
Decay~Modes ($\Delta S =0$)
 &\cite{Chang:2001pm} ($10^{-4}$)
 &\cite{Kiselev:2001zb} ($10^{-2}$)
 &\cite{Ivanov:2006ni} ($10^{-3}$)
 &\cite{Hernandez:2006gt} ($10^{-4}$)
 &\cite{Ebert:2010zu} ($10^{-4}$)
 &\cite{Wang:2011jt} ($10^{-4}$)
 &\cite{Rui:2017pre} ($10^{-3}$)
 &\cite{Lu:2025bvi} ($10^{-3}$)
 &\cite{Zhang:2023ypl} ($10^{-3}$)
 \\
\hline
${\rm BR}(B_c^+ \to \chi_{c0} \rho^+)$
& $7.2 $
& $3.3 $
& $1.3 $
& $6.7 $
& $5.8 $
& $7.6 $
& $ 5.8^{+1.3}_{-1.4} $
& $5.16^{+0.69}_{-0.87}(0.64^{+0.11}_{-0.13})$
& $ 1.69^{+0.20}_{-0.19} $
\\
\hline
${\rm BR}(B_c^+ \to \chi_{c1} \rho^+)$
& $2.9 $
& $0.46 $
& $0.29 $
& $1.0 $
& $1.5 $
& $2.3 $
& $ 2.8^{+0.5}_{-0.5} $
& $0.52^{+0.08}_{-0.13}(0.068^{+0.018}_{-0.018})$
& $ 0.43^{+0.01}_{-0.01} $
\\
\hline
${\rm BR}(B_c^+ \to \chi_{c2} \rho^+)$
& $5.1 $
& $3.2 $
& $1.2 $
& $6.5 $
& $11 $
& $5.6 $
& $ 16^{+4}_{-3} $
& $35.02^{+7.56}_{-1.88}(4.47^{+0.98}_{-0.35})$
&
\\
\hline \hline
Decay~Modes ($\Delta S=1$)
 &\cite{Chang:2001pm} ($10^{-6}$)
 &\cite{Kiselev:2001zb}
 &\cite{Ivanov:2006ni} ($10^{-5}$)
 &\cite{Hernandez:2006gt} ($10^{-5}$)
 &\cite{Ebert:2010zu} ($10^{-5}$)
 &\cite{Wang:2011jt} ($10^{-5}$)
 &\cite{Rui:2017pre} ($10^{-4}$)
 &\cite{Lu:2025bvi} ($10^{-3}$)
 &\cite{Zhang:2023ypl} ($10^{-5}$)
 \\
\hline
${\rm BR}(B_c^+ \to \chi_{c0} K^{*+})$
& $3.9 $
& $$
& $7.0 $
& $3.7 $
& $4.0 $
& $4.5 $
& $ 3.3^{+0.8}_{-0.7} $
& $0.27^{+0.04}_{-0.04}(0.033^{+0.006}_{-0.006})$
& $9.6^{+1.1}_{-1.1}$
\\
\hline
${\rm BR}(B_c^+ \to \chi_{c1} K^{*+})$
& $1.8 $
& $$
& $1.8 $
& $0.73 $
& $1.0 $
& $1.7 $
& $ 1.8^{+0.5}_{-0.3} $
& $0.032^{+0.005}_{-0.008}(0.0041^{+0.0011}_{-0.0010})$
& $2.7^{+0.1}_{-0.1}$
\\
\hline
${\rm BR}(B_c^+ \to \chi_{c2} K^{*+})$
& $3.1 $
& $$
& $6.5 $
& $3.8 $
& $7.4 $
& $3.3 $
& $ 9.6^{+2.2}_{-2.0} $
& $1.85^{+0.39}_{-0.10} (0.24^{+0.05}_{-0.02})$
&
\\
\hline \hline
\end{tabular}}
\end{center}
\end{table}

Next we inspect the contributions to the $B_c^+ \to \chi_{c1,c2} V^+$ decays
from the three polarization configurations, whose transversity amplitudes are written as
\beq
{\cal A}_{L}&=& \xi m^{2}_{B_c} A^{L},
\qquad
{\cal A}_{\parallel}=\xi \sqrt{2}m^{2}_{B_c} A^{N},
\qquad
{\cal A}_{\perp}=\xi m_{V} m_{\chi_{c1,c2}}
\sqrt{2(r^{2}-1)} A^{T}\;,
\label{eq:ase}
\eeq
with the normalization factor $\xi=\sqrt{G^2_{F}|{\bf P_c}|/(16\pi m^2_{B_c}\Gamma)}$
and the ratio $r=P_{2}\cdot P_{3}/(m_{V}\cdot m_{\chi_{c1,c2}})$.
For their corresponding factorization formulas, refer to~\cite{Rui:2017pre}.
These transversity amplitudes obey the normalization
\beq
|{\cal A}_{L}|^2+|{\cal A}_{\parallel}|^2+|{\cal A}_{\perp}|^2 &=& 1 \;.
\eeq
The {\it CP}-averaged longitudinal polarization fraction $f_{L}$ is defined as
\beq
f_{L}&\equiv&
\frac{|{\cal A}_{L}|^2}
{|{\cal A}_L|^2+|{\cal A}_{||}|^2+|{\cal
A}_{\perp}|^2}\;.
\label{eq:pf}
\eeq

We predict the $B_c^+ \to \chi_{c1,c2} V^+$ BRs
\beq
{\rm BR}(B_c^+ \to \chi_{c1} \rho^+)&=&
1.40
^{+0.49}_{-0.34}(\beta_{B_c})
^{+0.22}_{-0.21}(f_{c\bar c})
^{+0.00}_{-0.00}(a_\rho)
^{+0.06}_{-0.05}(V_{cb})\times 10^{-3} \;,
\label{eq:c1ro}
\\
{\rm BR}(B_c^+ \to \chi_{c1} K^{*+}) &=&
2.49
^{+0.85}_{-0.57}(\beta_{B_c})
^{+0.41}_{-0.36}(f_{c\bar c})
^{+0.05}_{-0.05}(a_{K^*})
^{+0.10}_{-0.09}(V_{cb})\times 10^{-5}\;,
\\
{\rm BR}(B_c^+ \to \chi_{c2} \rho^+)&=&
1.18
^{+0.36}_{-0.26}(\beta_{B_c})
^{+0.27}_{-0.24}(f_{c\bar c})
^{+0.01}_{-0.00}(a_\rho)
^{+0.05}_{-0.04}(V_{cb})\times 10^{-3}  \;,
\label{eq:c2ro}
\\
{\rm BR}(B_c^+ \to \chi_{c2} K^{*+})&=&
6.83
^{+2.05}_{-1.49}(\beta_{B_c})
^{+1.58}_{-1.42}(f_{c\bar c})
^{+0.09}_{-0.09}(a_{K^*})
^{+0.28}_{-0.24}(V_{cb})\times 10^{-5} \;,
\eeq
where the uncertainties are mainly caused by the variations of the $B_c$ meson shape parameter
$\beta_{B_c}$ and the charmonium decay constants. The large
$B_c^+ \to \chi_{c1,c2} \rho^+$ BRs are measurable at the LHC.
We also predict the ratio between the $B_c^+ \to \chi_{c1} \rho^+$ and $B_c^+ \to J/\psi \pi^+$
BRs,
\beq
R^{\chi_{c1}\rho}_{J/\psi \pi} &\equiv& \frac{{\rm BR}(B_c^+ \to \chi_{c1}\rho^+)}{{\rm BR}(B_c^+ \to J/\psi \pi^+)}=1.20^{+0.13}_{-0.12}\;,
\label{eq:r-c1rojpi}
\eeq
and the ratio between the $B_c^+ \to \chi_{c2} \rho^+$
and $B_c^+ \to J/\psi \pi^+$ BRs,
\beq
R^{\chi_{c2}\rho}_{J/\psi \pi} &\equiv& \frac{{\rm BR}(B_c^+ \to \chi_{c2}\rho^+)}{{\rm BR}(B_c^+ \to J/\psi \pi^+)}=1.01^{+0.15}_{-0.15}\;.
\label{eq:r-c2rojpi}
\eeq
Compared with the results in Table~\ref{tab:BRs-chicjV}, our
${\rm BR}(B_c^+ \to \chi_{c1} \rho^+)$ is similar to that in Ref.~\cite{Rui:2017pre}, and
our ${\rm BR}(B_c^+ \to \chi_{c1} K^{*+})$ and ${\rm BR}(B_c^+ \to \chi_{c2} V^+)$ agree
well with those
in Refs.~\cite{Lu:2025bvi,Zhang:2023ypl} and \cite{Ivanov:2006ni,Ebert:2010zu}, respectively.
All the above predictions in the iPQCD formalism can be tested in future experiments.

The longitudinal polarization fractions associated with the $B_c^+ \to \chi_{c1,c2} V^+$ decays
are given by
\beq
f_L(B_c^+ \to \chi_{c1} \rho^+)&=&
(95.0^{+0.9}_{-1.1})
\%  \;,
\qquad
f_L(B_c^+ \to \chi_{c1} K^{*+})=
(79.7^{+3.7}_{-3.4})
\% \;,\\
f_L(B_c^+ \to \chi_{c2} \rho^+)&=&
(92.8^{+1.1}_{-1.3})
\%  \;,
\qquad
f_L(B_c^+ \to \chi_{c2} K^{*+})=
(90.9^{+1.4}_{-1.6})
\% \;,
\eeq
which indicate the dominance of the longitudinal components in these
modes. Our $f_L(B_c^+ \to \chi_{c1} \rho^+)$ and $f_L(B_c^+ \to \chi_{c1} K^{*+})$ differ
slightly from those in the conventional PQCD approach~\cite{Rui:2017pre}, while
$f_L(B_c^+ \to \chi_{c2} V^+)$ are consistent.

The ratios between the $B_c^+ \to \chi_{cJ} (P, V)^+$ BRs read, for
$B_c^+ \to \chi_{c1,c2} K^{*+}$ and $B_c^+ \to \chi_{c1,c2} \rho^+$,
\beq
R_{K^*/\rho}^{\chi_{c1}} &\equiv&
\frac{{\rm BR}(B_c^+ \to \chi_{c1} K^{*+})}{{\rm BR}(B_c^+ \to \chi_{c1} \rho^+)} =
(1.78^{+0.04}_{-0.04})  \times 10^{-2}
\;,
\label{eq:r-rhoks-c1}\\
R_{K^*/\rho}^{\chi_{c2}}&\equiv&
\frac{{\rm BR}(B_c^+ \to \chi_{c2} K^{*+})}{{\rm BR}(B_c^+ \to \chi_{c2} \rho^+)}
=  (5.79^{+0.03}_{-0.09}) \times 10^{-2}
\;.
\label{eq:r-rhoks-c2}
\eeq
The latter is close to $R_{K^*/\rho}^{J/\psi}$ in~\cite{Liu:2023kxr}, but the former is not.
This disparity traces back to the behaviors of the charmonium distribution amplitudes
in Eqs.~(\ref{eq:c1-t2-L})$-$(\ref{eq:c1-t3-v}) and~(\ref{eq:c2-t2-L})$-$(\ref{eq:c2-t3-v}):
the overall antisymmetric $\chi_{c2}$ distribution amplitudes result in
the analogy to the decays $B_c^+ \to J/\psi (\rho, K^*)^+$, i.e., the similarity between
$R_{K^*/\rho}^{\chi_{c2}}$ and $R_{K^*/\rho}^{J/\psi}$.
The $\chi_{c1}$ distribution amplitudes are (anti)symmetric for the (transverse)
longitudinal polarization at leading twist, and (anti)symmetric for the (longitudinal)
transverse polarization at twist 3. These properties, producing distinct nonfactorizable
emission contributions, differentiate the $B_c^+ \to \chi_{c1} (\rho, K^*)^+$
and $B_c^+ \to J/\psi (\rho, K^*)^+$ channels.

The combinations of the $B_c^+ \to \chi_{cJ} (\pi^+, \rho^+)$ BRs in Eqs.~(\ref{eq:c2pi}),
(\ref{eq:c0pi}), (\ref{eq:c1pi}), (\ref{eq:c0ro}), (\ref{eq:c1ro}) and (\ref{eq:c2ro})
give rise to the ratios
\beq
R_{\rho/\pi}^{\chi_{c0}} &\equiv&
\frac{{\rm BR}(B_c^+ \to \chi_{c0} \rho^+)}{{\rm BR}(B_c^+ \to \chi_{c0} \pi^+)} =
1.47^{+0.06}_{-0.09}\;,
\\
R_{\rho/\pi}^{\chi_{c1}} &\equiv&
\frac{{\rm BR}(B_c^+ \to \chi_{c1} \rho^+)}{{\rm BR}(B_c^+ \to \chi_{c1} \pi^+)} =
33.65^{+7.70}_{-5.28}\;.
\\
R_{\rho/\pi}^{\chi_{c2}} &\equiv&
\frac{{\rm BR}(B_c^+ \to \chi_{c2} \rho^+)}{{\rm BR}(B_c^+ \to \chi_{c2} \pi^+)} =
3.15^{+0.06}_{-0.05}\;,
\eeq
in a nonuniform pattern. The value of $R_{\rho/\pi}^{\chi_{c2}}$ matches perfectly the
ratio $3.15^{+0.09}_{-0.10}$ between the $B_c^+ \to J/\psi \rho^+$ and
$B_c^+ \to J/\psi \pi^+$ BRs~\cite{Liu:2023kxr}. The value of $R_{\rho/\pi}^{\chi_{c1}}$,
in spite of the significant theoretical uncertainties, is surprisingly huge and marks a
sharp contrast to $R_{\rho/\pi}^{J/\psi}$. A careful check unveils that the contributions
from the twist-2 and twist-3 $\chi_{c1}$ meson distribution amplitudes are constructive
(destructive) in the $B_c^+\to \chi_{c1} \rho^+$ ($B_c^+\to \chi_{c1} \pi^+$) mode.
Such peculiar interferences do not appear in the $\chi_{c0,c2}$ involved decays. It is
encouraged to examine the iPQCD predictions and to deepen our understanding of the
intricate dynamics in $B_c\to\chi_{cJ}$ transitions by measuring the above ratios.

At last, adopting the BRs of the $\chi_{cJ}$ strong decays in
Eqs.~(\ref{eq:c0-ppkk})$-$(\ref{eq:c2-ppkk}) and (\ref{eq:c0-2ppkk})$-$(\ref{eq:c2-2ppkk}),
we get the BRs of $B_c^+ \to \chi_{cJ} \rho^+$ with secondary decay chains under the
narrow-width approximation,
\beq
{\rm BR}(B_c^+ \to  \chi_{c0} (\to \pi^+ \pi^-)\rho^+)&\equiv&
{\rm BR}(B_c^+ \to  \chi_{c0} \rho^+)\cdot {\rm BR}( \chi_{c0}\to \pi^+ \pi^-)
= (1.80^{+0.69}_{-0.54}) \times 10^{-6},
\\
{\rm BR}(B_c^+ \to \chi_{c0} (\to K^+ K^-) \rho^+)&\equiv&
{\rm BR}(B_c^+ \to \chi_{c0} \rho^+) \cdot {\rm BR}(\chi_{c0} \to K^+ K^-)
= (1.89^{+0.72}_{-0.56}) \times 10^{-6},
\\
{\rm BR}(B_c^+ \to \chi_{c2} (\to \pi^+ \pi^-) \rho^+)&\equiv&
{\rm BR}(B_c^+ \to \chi_{c2} \rho^+)\cdot {\rm BR}(\chi_{c2} \to \pi^+ \pi^-)
= (1.90^{+0.74}_{-0.58}) \times 10^{-6},
\\
{\rm BR}(B_c^+ \to \chi_{c2} (\to K^+ K^-) \rho^+)&\equiv&
{\rm BR}(B_c^+ \to \chi_{c2} \rho^+)\cdot {\rm BR}(\chi_{c2} \to K^+ K^-)
= (1.44^{+0.56}_{-0.44}) \times 10^{-6},
\eeq
and
\beq
{\rm BR}(B_c^+ \to \chi_{c0} (\to \pi^+ \pi^- \pi^+ \pi^-) \rho^+)&\equiv&
{\rm BR}(B_c^+ \to \chi_{c0} \rho^+) \cdot {\rm BR}(\chi_{c0} \to 2(\pi^+ \pi^-))
\non
&=& (6.32^{+2.42}_{-1.90}) \times 10^{-6}\;,
\\
{\rm BR}(B_c^+ \to \chi_{c0} (\to \pi^+ \pi^-K^+ K^-) \rho^+ )&\equiv&
{\rm BR}(B_c^+ \to \chi_{c0} \rho^+)\cdot {\rm BR}(\chi_{c0} \to \pi^+ \pi^-K^+ K^-)
\non
&=& (5.41^{+2.11}_{-1.67}) \times 10^{-6}\;,
\\
{\rm BR}(B_c^+ \to \chi_{c1} (\to \pi^+\pi^-\pi^+ \pi^-) \rho^+)&\equiv&
{\rm BR}(B_c^+ \to \chi_{c1} \rho^+)\cdot {\rm BR}(\chi_{c1} \to 2(\pi^+\pi^-))
\non
&=& (0.96^{+0.37}_{-0.28}) \times 10^{-5}\;,
\\
{\rm BR}(B_c^+ \to \chi_{c1} [\to \pi^+ \pi^-K^+ K^-] \rho^+)&\equiv&
{\rm BR}(B_c^+ \to \chi_{c1} \rho^+)\cdot {\rm BR}(\chi_{c1} \to \pi^+ \pi^-K^+ K^-)
\non
&=& (6.30^{+2.80}_{-2.28}) \times 10^{-6}\;,
\\
{\rm BR}(B_c^+ \to \chi_{c2} (\to \pi^+\pi^-\pi^+ \pi^-) \rho^+)&\equiv&
{\rm BR}(B_c^+ \to \chi_{c2} \rho^+)\cdot {\rm BR}(\chi_{c2} \to 2(\pi^+ \pi^-))
\non
&=& (1.36^{+0.54}_{-0.42}) \times 10^{-5}\;,
\\
{\rm BR}(B_c^+ \to \chi_{c2} (\to \pi^+ \pi^-K^+ K^-) \rho^+)&\equiv&
{\rm BR}(B_c^+ \to \chi_{c2} \rho^+)\cdot {\rm BR}(\chi_{c2} \to \pi^+ \pi^-K^+ K^-)
\non
&=& (0.98^{+0.40}_{-0.33}) \times 10^{-5}\;.
\eeq
These predictions provide a useful reference for experimental search for multibody
$B_c$ meson decays.

\section{Summary} \label{sec:summary}

We have studied the $B_c$-meson decays into $\chi_{cJ}$ plus light pseudoscalar or vector
mesons in the leading-order iPQCD approach. The $B_c^+ \to \chi_{cJ} (P, V)^+$ BRs and the
longitudinal polarization fractions of the $B_c^+ \to \chi_{c1,c2} V^+$ decays were evaluated.
In view of the unclear $B_c$ meson production cross sections, we defined and predicted
the ratios between the $B_c^+ \to \chi_{cJ} (P, V)^+$ and $B_c^+ \to J/\psi \pi^+$ BRs
for experimental verifications. In summary, we found that
the predicted ratios $R_{\chi_{c2}/J/\psi} = 0.32 \pm 0.05$ and
$R_{\chi_{c1}/\chi_{c2}} = 0.11 \pm 0.01$ coincide with the current
observations and upper limit reported by the LHCb,
and the ratios $R_{\chi_{c0}/J/\psi} = 0.17 \pm 0.02$ and
$R_{\chi_{c1}/J/\psi} = (3.56^{+0.81}_{-0.82})  \times 10^{-2}$ can be confronted by
future measurements. The different ratios $R_{\rho/\pi}^{\chi_{cJ}}$ were obtained:
$R_{\rho/\pi}^{\chi_{c2}} = 3.15^{+0.06}_{-0.05}$ matches perfectly
$R_{\rho/\pi}^{J/\psi}$, but $R_{\rho/\pi}^{\chi_{c0}}=1.47^{+0.06}_{-0.09}$
$(R_{\rho/\pi}^{\chi_{c1}}=33.65^{+7.70}_{-5.28})$ is a bit lower (surprisingly higher)
than $R_{\rho/\pi}^{J/\psi}$. This nonuniform pattern is attributed to the distinct
interplays between the twist-2 and twist-3 contributions in these decays.
The predicted ratios $R_{K/\pi}^{\chi_{cJ}}$ and $R_{K^*/\rho}^{\chi_{c2}}$ meet the naive
expectation that they are governed by factorizable emission contributions. The larger (smaller)
$R_{K^*/\rho}^{\chi_{c0}} (R_{K^*/\rho}^{\chi_{c1}})$, being evidently disparate from the
naive anticipation, suggests the varying involved nonfactorizable emission contributions.
Inputting the measured ${\rm BR}(\chi_{c0,c2} \to \pi^+\pi^-/K^+K^-)$ and
${\rm BR}(\chi_{cJ} \to \pi^+\pi^-(\pi^+\pi^+/K^+K^-))$, the BRs of
$B_c^+ \to \chi_{cJ} (\pi, \rho)^+$ with secondary decay chains
were estimated under the narrow-width approximation, which serve as useful references for
their experimental search.


\begin{acknowledgments}
X.L. thanks Vanya~Belyaev, Hai-Yang~Cheng, Pei-Lian~Li, M.A.~Olpak, Cong-Feng~Qiao,
Guo-Li~Wang, Wei~Wang, Yu-Ming~Wang, Jie-Sheng~Yu, Rui~Zhou,
and Rui-Lin~Zhu for their helpful discussions and correspondences.
This work is supported by the National Natural Science
Foundation of China under Grants No.~11875033 and No.~12335003,
and by the National Science
and Technology Council of the Republic of
China under Grant No. NSTC-113-2112-M-001-024-MY3.
\end{acknowledgments}


\end{CJK*}
\end{document}